\documentclass[twocolumn]{revtex4-1}%
\usepackage{graphicx}
\usepackage{float}
\usepackage{dcolumn}
\usepackage{bm}
\usepackage{amsmath}
\usepackage{amsfonts}
\usepackage{amssymb}%
\usepackage{epstopdf}
\usepackage{epsfig}
\begin{document}

\title{Laser Chimeras as a paradigm for multi-stable patterns in
  complex systems}

\author{Laurent Larger$^{\dag}$, Bogdan Penkovsky$^\dag$, Yuri
  Maistrenko$^{\dag\ddag}$ } \affiliation{$\dag$FEMTO-ST / Optics
  Dept., UMR CNRS 6174, University of Franche-Comt\'{e}, 15 avenue des
  Montboucons, 25030 Besan\c{c}on Cedex, France} \affiliation{$\ddag$
  Institute of Mathematics and Center for Medical and Biotechnical
  Research, NAS of Ukraine, Tereschenkivska Str. 3, 01601 Kyiv,
  Ukraine}

\date{\today }

\begin{abstract}
  Chimera is a rich and fascinating class of self-organized solutions
  developed in high dimensional networks having non-local and symmetry
  breaking coupling features. Its accurate understanding is expected
  to bring important insight in many phenomena observed in complex
  spatio-temporal dynamics, from living systems, brain operation
  principles, and even turbulence in hydrodynamics. In this article we
  report on a powerful and highly controllable experiment based on
  optoelectronic delayed feedback applied to a wavelength tunable
  semiconductor laser, with which a wide variety of Chimera patterns
  can be accurately investigated and interpreted. We uncover a cascade
  of higher order Chimeras as a pattern transition from $N$ to $N-1$
  clusters of chaoticity.  Finally, we follow visually, as the gain
  increases, how Chimera is gradually destroyed on the way to apparent
  turbulence-like system behaviour.
\end{abstract}


\maketitle

\section*{Paradigm for complexity: space-time vs. time delay?}

Complexity usually develops in high dimensional systems involving
nonlinear interactions between system variables. Straightforward
paradigmatic experiments to explore and to understand complexity, are
generally thought as spatio-temporal nonlinear dynamics with obvious
infinite number of degrees of freedom.  Such systems provide solutions
in high- or infinite dimensional phase space, which degree of
complexity depends on the strength of nonlinear effects, as well as on
the distribution of coupling between the phase space coordinates
(network nodes). The equations of motion appear as a mathematical
translation of the deterministic origin ruling the dynamics,
e.g. Navier-Stokes equations in fluid dynamics, Ginzburg-Landau
equation from superconducting phase transition, reaction-diffusion
systems of biological relevance, nonlinear Schr\"odinger equation in
nonlinear optics, and so on. These generic models come however with
difficult theoretical analysis as they are essentially nonlinear, and
many unsolved hard problems still remain.

Understanding the rules governing complexity remains a Human
challenge, since this is our own natural environment, from living
systems to society. But do we necessarily need complicated equations
to understand diversity in the Nature?  For some motions like chaos,
simple equations have been identified, such as the 1D logistic
(quadratic) map or the 3D Lorenz system, from which complex dynamical
mechanisms, among which is chaos, have been clearly identified and
understood.

\begin{figure}[ht]
  \begin{center}
    \includegraphics[width=0.45\textwidth]{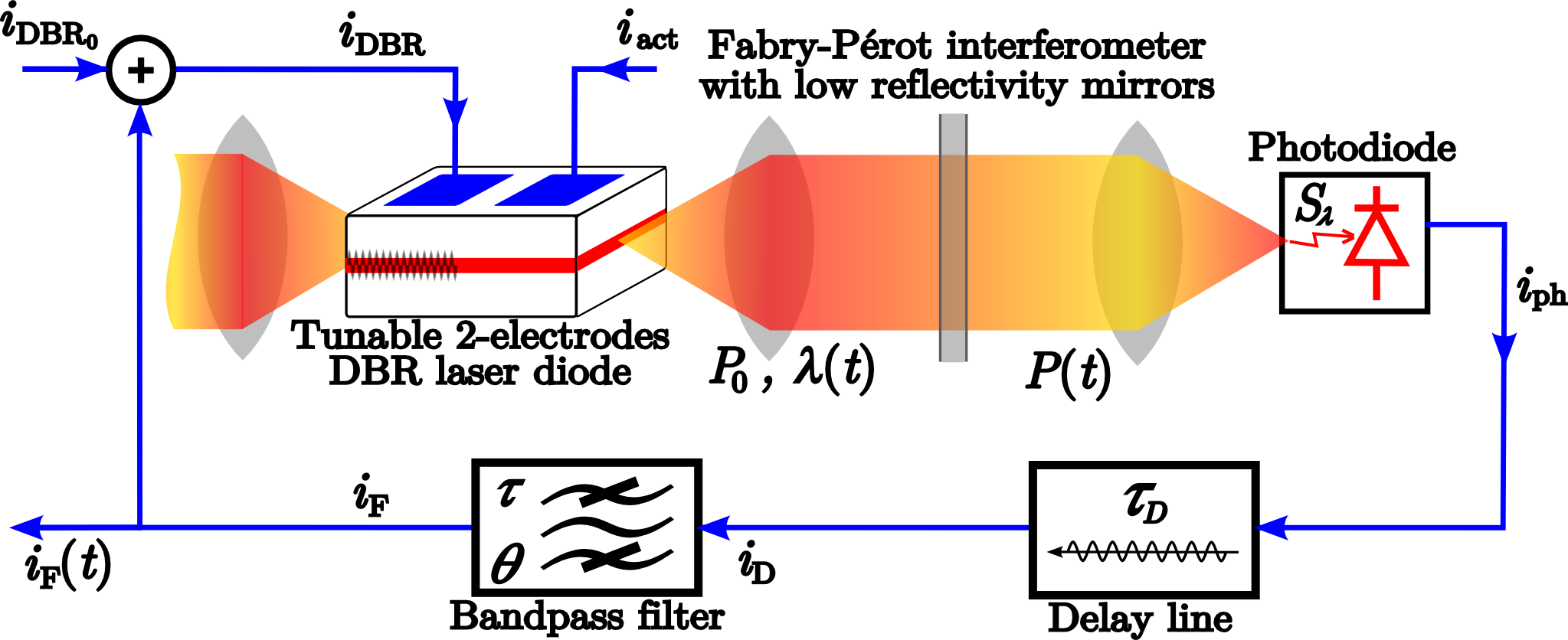}
    \caption{Tunable semiconductor laser setup allowing for highly
      controllable multiple head Chimera
      states. \label{fig:lambda_setup}}
  \end{center}
\end{figure}

Beyond the inspiring achievements in nonlinear dynamics theory, one of
the nowadays hot research topics aimed at understanding complex
motions and systems, is related to deterministic organisation of
networks of oscillators in both finite- and infinite-dimensional
situation. Within this topic, a particular phenomenon discovered in
the early 2000, have attracted a strongly growing attention, namely
{\em Chimera state}. Such solutions have been defined as the emergence
of ``incongruent'' patterns of co-existing synchronous and incoherent
behaviours, where different groups of oscillators within the network
are exhibiting motions that are similar within a cluster, but
``incongruent'' between clusters. Discovered in 2002 by Kuramoto and
Battogtokh \cite{kuramoto:npcs02}, Chimeras states \cite{abrams:prl04}
have been experimentally observed in 2012 only, on spatio-temporal
dynamics in the transverse plane of a light beam
\cite{hagerstrom:nphy12}, and independently in the volume of a
chemical reaction \cite{tinsley:nphy12} and soon after, in mechanical
experiments with coupled metronomes
\cite{martens:pnas13,kapitaniak:sr14}.  Recently, Chimera motions were
revealed in an even simpler system, however known for its infinite
dimensional phase space, a nonlinear electronic delayed self feedback
oscillator \cite{larger:prl13}. Exploring further this unexpected
feature for the class of delay dynamics, we have specially designed a
laser-based setup in which a complex organization of many multistable
Chimera states can be obtained and described in a detailed way.  Most
importantly, thanks to the high control accuracy of the optoelectronic
setup, novel Chimera features have been identified experimentally, and
have been found in excellent qualitative agreement with numerical
simulations, which will potentially open new fundamental as well as
applied perspectives for Chimera states.


The article is organized as follows. We will first detail the
structural and mathematical requirements of a delay dynamical system
in order to obtain Chimera, which requirements have closely guided the
design of the experiment based on a tunable laser diode. The laser
setup will then be presented in section \ref{sec:setup}. Section
\ref{sec:eps_del} will report on the discovery of Chimera order
cascade in a 2D-parameter space (spatio-temporal coupling parameters),
as well as on the transition to turbulent-like behaviour when moving
along a third parameter (feedback gain). The conclusion will propose
to extend implications of our findings with delay systems onto other
fields concerned by complex nonlinear dynamics, such as photonic
brain-inspired computing and turbulent-laminar processes in fluid
mechanics.

%
\section{Delay dynamics requirements for Chimera}

A delay differential equation (DDE) has the specificity to be a purely
temporal dynamics, however having an infinite dimensional phase space
\cite{mackey:sci77, ikeda:oc79, chow:sv87} as for spatio-temporal
dynamics modeled by partial differential equations. The common scalar
form of such a DDE is often given by an equation
$\varepsilon\dot{x}(s)=-x(s)+f[x(s-1)]$, where $s$ is the time
variable normalized to the delay, and $f$ represents a nonlinear
transformation of the amplitude variable $x$.  Though being scalar and
of a first order only, the initial conditions required to uniquely
define a solution, take the form of a functional of time defined over
a time delay interval i.e. $x_0(s)$ s.t. $s\in[-1;0]$. In the case of
a large physical delay, which implies $\varepsilon\ll1$, the strong
multiple time scale character allows for high complexity phase space
\cite{leberre:pra90}, and spatio-temporal analogy have been proposed
already more than 20 years ago \cite{arecchi:pra92} to help
understanding the underlying system behaviour. The simple idea behind
such space-time analogy is to represent the dynamics as the evolution
with the discrete time $n$ of the functional trajectory $\{x_n(\sigma)
= x(s)\, | \, s = \sigma + n\eta, \, \text{with} \,
\sigma\in[0;\eta]\, \text{and} \, n\in\mathbb{N}\}$. This results in a
1-Dimensional ``spatial'' distribution of continuously coupled
amplitudes, over a finite ``virtual space'' interval ($[0;\eta]$, with
$\eta=1+\gamma$ and $\gamma = \text{O}(\varepsilon)$). The dynamics of
this functional trajectory appear as a discrete iteration from $n$ to
$(n+1)$ according to a time step $\eta$ close to unity (i.e. the
delay). Space refers thus to the fast time scale $\sigma\in[0;\eta]$
(spatial ``granularity'' being of the order of $\varepsilon$), whereas
discrete time variable $n$ refers to the long time scale counting
roughly the
number of time delays.\\

In delay systems, Chimera is thus expected to appear as a rich
multi-clustered pattern $x_n(s)$, self-sustained as $n$ is
growing. However, in its simplest scalar form a DDE does not allow for
such (nearly) one-delay periodic patterns as they are Lyapunov
unstable at any $\varepsilon>0$ \cite{sharkovsky93}, all the space
being rapidly filled by a unipolar amplitude
\cite{giacomelli:epl12}. We recently showed \cite{larger:prl13} that
introducing a slow integral term to the DDE allows for the
stabilization of such patterns, giving rise to robust Chimera:
\begin{equation}
  \varepsilon \frac{\text{d}x}{\text{d}s}(s) + x(s) + \delta\int_{s_0}^s
  x(\xi)\,\text{d}\xi = f[x(s-1)].\label{eq:idde}
\end{equation}
Additional requirement for obtaining Chimera concerns the function
$f$, which associated map ($x_{n+1}=f[x_n]$) has to exhibit
multi-stability through a positive feedback at the zero unstable
operating point, connecting two asymmetric extrema (a broad minimum
for $x<0$ leading to a stable equilibrium, and a sharp
maximum for $x>0$ leading to an unstable fixed point and chaos around it).\\
Rewriting the integro-differential delay equation into a more common
form, one obtains the following two coupled first order delay
equation:
\begin{align}
  \varepsilon x^{\prime} & =-\delta\,y - x + f[x(s-1)], \nonumber\\
  y^{\prime} & =x,\label{eq:idde_norm}
\end{align}
where the additional slow variable $y$ accounts for the added integral
term which weight is controlled by the additional parameter
$\delta>0$. This formulation allows to get a qualitative but effective
analysis of the dynamics in terms of simplified slow-fast
two-dimensional dynamics \cite{weicker:ptrsA13,larger:prl13}, it
however does not remove the difficulty to interpret detailed
microscopic nature of the spatio-temporal phenomenon. Another
formulation of such dynamics, more inspired by signal theory, involves
a simple convolution product with the so-called filter impulse
response $h(s)$, $x(s)=\int h(s-\xi)f[x(\xi-1)]\,\text{d}\xi$. $h(s)$
is originating from the linear left hand side of Eq.(\ref{eq:idde}),
its Fourier transform being simply the corresponding linear Fourier
frequency filtering function. The nonlinear delay dynamics is thus
revealed as a feedback loop oscillator as in
Fig.\ref{fig:lambda_setup}, in which a linear filter provides the
argument $x$ for the nonlinear function $f$, which output is delayed
and serves then as the filter input. This ``convolution product''
allows for a straightforward space-time analogy, since rewriting it in
order to reveal the discrete functional trajectory $x_n(\sigma)$, one
obtains (see Supplementary Material):
\begin{equation}
  \label{eq:conv_dde}
  x_n(\sigma) = x_{n-1}(\sigma) + \int_{\sigma-1}^{\sigma+\gamma}
  h(\sigma+\gamma-\xi) \cdot f[x_{n-1}(\xi)]\,\text{d}\xi,
\end{equation}
which shows the mapping dynamics from $x_{n-1}$ to $x_n$, via a
spatial nonlinear and non-local coupling between the position $\sigma$
and a $\xi-$shift around it, with
$\xi\in[\sigma-\Delta;\sigma+\gamma]$. Equation (\ref{eq:conv_dde})
thus reveals $f$ as the nonlinear coupling function, $h$ as the
coupling weight, and the quantity $(\Delta +\gamma)$ is an effective
spatial range for the coupling, which spread is practically determined
by the width of the impulse response $h$ (see Supplementary Material).

\section{Laser wavelength delay dynamics}\label{sec:setup}
The photonic setup is depicted in Fig.\ref{fig:lambda_setup}. It is
designed according to the above described requirements allowing for
the observation of Chimera in a delay dynamics. The physics and
photonic concepts are inspired by a wavelength chaos generator
designed previously for optical chaos communication, where chaos was
obtained from an Ikeda dynamics (given by DDE with a symmetric
nonlinear function $f[x]=\beta\sin^2(x+\Phi_0)$ \cite{larger:jqe98}).
Two essential modifications are allowing for Chimera: (\textit{i}) a
bandpass filter instead of low pass one is providing the left hand
side in Eq.(\ref{eq:idde}), and (\textit{ii}) a Fabry-P\'erot (FP)
interferometer instead of a birefringent one which is providing the
asymmetric nonlinear Airy function:
\begin{equation}
  \label{eq:airy}
  f[x]=\frac{\beta}{1+m\sin^2(x+\Phi_0)}.
\end{equation}

The oscillating principles of such an optoelectronic tunable laser
delay oscillator are as follow. A dual electrode tunable distributed
Bragg reflector (DBR) laser diode emitting at 1.5~$\mu$m provides a
laser beam, which wavelength deviation (corresponding to variable $x$)
is proportional to a tuning electrode current $i_\text{DBR}$ (another
conventional active electrode receives the injection current
$i_\text{act}$ setting the emitted optical power). An offset current
$i_{\text{DBR}_0}$ is added to the tuning electrode for the central
laser wavelength, thus allowing for the setting of the appropriate
parameter $\Phi_0$ in Eq.(\ref{eq:airy}) (needed for the positive
feedback condition). Wavelength fluctuations are then non linearly
converted into intensity ones through the FP, as $x$ is scanning back
and forth the Airy function from the flat destructive interference
condition up to the sharp constructive one. A photodiode is used to
convert the optical intensity fluctuations into an electrical signal,
which is then delayed in time by $\tau_D$ thanks to an easily tunable
electronic delay line. This electronic path allows for an accurate
control of the equation of motion (\ref{eq:idde}), through an
appropriate bandpass filter having characteristic times $\theta$ and
$\tau$ defining the low and high cut-off frequencies of the filter
respectively. The signal is finally amplified (setting the normalized
gain $\beta$ in Eq.(\ref{eq:airy})), before being fed back
($i_\text{F}$) onto the laser DBR tuning electrode. The electronic
filter output $i_\text{F}$ is the monitored time trace proportional to
$x(s)$, from which space-time Chimera patterns
$\{x_n(\sigma)\,|\,\sigma\in[0;\eta],\,n\in\mathbb{N}\}$ can be
extracted and analyzed depending on the system parameters. The most
important normalized parameters are the small quantities
$\varepsilon=\tau/\tau_D$ and $\delta=\tau_D/\theta$ introduced in
Eq.(\ref{eq:idde}). These two quantities are practically influencing
the actual shape and ``spatial'' spread of the impulse response $h$,
$(\gamma+\Delta)\simeq\varepsilon\ln(\varepsilon\delta)^{-1}$, as
discussed for Eq.(\ref{eq:conv_dde}) for the analogy of a weighting
function in a network of non-locally coupled oscillators.

\begin{figure}[ht]
  \begin{center}
    \includegraphics[width=0.50\textwidth]{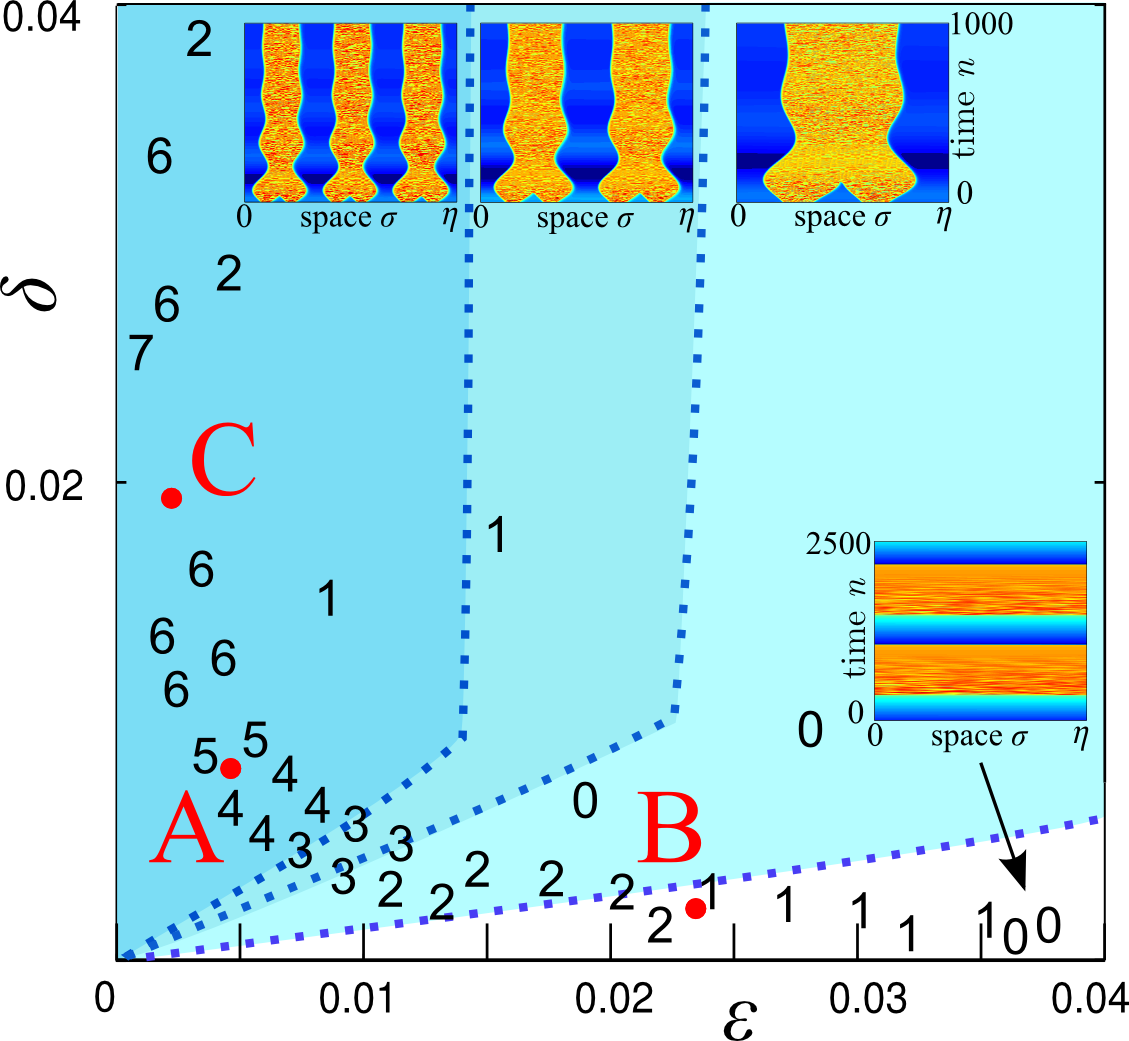}
    \caption{(a) Observed numerical (colorized) and experimental
      (integer $N$) solutions in the parameter plane
      $(\varepsilon,\delta)$, with $\beta=2.0$ and
      $\Phi_0=-0.4$. $N=0$ or white region stands for chaotic breather
      \cite{kouomou:prl05}, and otherwise one has $N-$headed
      chimeras. Crossing a dotted line (numerically determined) from
      the up-left to the low-right means a unit decrement for the
      maximal number $N$ of observable chimera heads (any $N_\sigma$
      heads with $N_\sigma\le N$ being possible). Insets are example
      of space-time plots for $N=1,2,3$ (Chimeras) and 0 (chaotic
      breather).\label{fig:eps_del}}
  \end{center}
\end{figure}

\section{Cascade of multi-headed Chimera}\label{sec:eps_del}

\begin{figure}[ht]
  \begin{center}
    \includegraphics[width=0.45\textwidth]{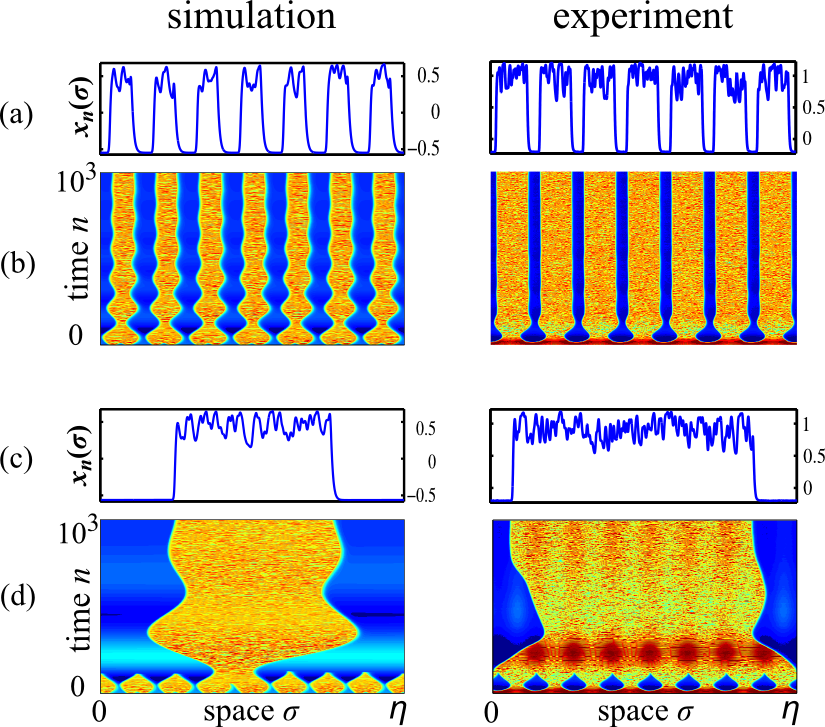}
    \caption{Examples of emergence and stabilization of $N_\sigma
      -$headed Chimera at point C in Fig.\ref{fig:eps_del}
      ($N_\sigma=7$ and 1 for (a,b) and (c,d) respectively). Left
      panel: Numerics; Right panel: Experiment. (a,c): asymptotic
      (biggest $n$) functional trajectory $x_n(\sigma)$. (b,d)
      spatio-temporal pattern birth of $N_\sigma -$headed
      Chimera. (b): 7-periodic small initial forcing; (d): same as (b)
      but with 8-periodic small initial forcing; $N_\sigma=8$ is not
      stable, thus leading to a single-headed asymptotic chimera
      state.}
    \label{fig:high_order_chim}
  \end{center}
\end{figure}

As reported in \cite{larger:prl13}, Chimera in DDE is revealed as the
spontaneous emergence of a particular functional pattern $x_n(\sigma)$
showing sub-intervals over $[0;\eta]$ each of which being
characterized whether by a nearly constant negative amplitude, or by a
chaotic-like oscillations (see space-time patterns in the in-box of
Fig.\ref{fig:eps_del}, and time traces in
Fig.\ref{fig:high_order_chim}). An amazing peculiarity is that such
rich and organized functional behaviour in $\sigma$ can be self
sustained as $n$ is iterated. When $N_\sigma$ chaotic intervals of
this kind exist for $\sigma \in [0;\eta]$, the Chimera is
referred as to a $N_\sigma-$\textit{headed Chimera} state.\\
Through the experimental behavior observed from the laser setup in
Fig.\ref{fig:lambda_setup}, as well as from the respective numerical
simulations of its established model in Eq.(\ref{eq:idde_norm}), the
$(\varepsilon,\delta)$-parameter space is discovered to contain a
specific multistable bifurcation structure. This structure reveals
cascaded regions from the low-right to the up-left of the
$(\varepsilon,\delta)-$plane, which are successively embedded one in
the other with increasing maximal integer $N$ of any $N_{\sigma}\le N$
number of possible Chimera heads.  As shown in Fig.\ref{fig:eps_del},
these regions are delimited by bifurcation curves characterized by the
transition from to $N$ to $(N-1)$ as $\varepsilon$ is increased, and
$\delta$ is decreased. These regions accumulate with increasing $N$
close to the $\varepsilon=0^+$-axis.  On the opposite side, the lowest
value $N=0$ finishes on the $\delta=0^+$-axis. This region does not
lead any stable Chimera motion, but reveals so-called \textit{chaotic
  breather} solutions \cite{kouomou:prl05}, a slow envelope
alternating fast chaotic oscillations and slow drifts, over time
durations of the order of $\delta^{-1}$. Typical Chimera patterns and
chaotic breather dynamics are shown in insets of
Fig. \ref{fig:eps_del}. Numerics (full scan) and experiments (for
which a few points only are explored while decreasing $\tau_D$ to scan
a hyperbola defined by $\varepsilon\delta=$constant) reveal excellent
qualitative agreement in the observation of this unusual bifurcation
structure. Quantitative discrepancies between numerics and experiments
are noticed for the absolute position of the $N-$transition lines,
probably related to the influence of noise as well as to uncertain
calibration of experimental parameters.

\begin{figure}[ht]
  \begin{center}
    \includegraphics[width=0.45\textwidth]{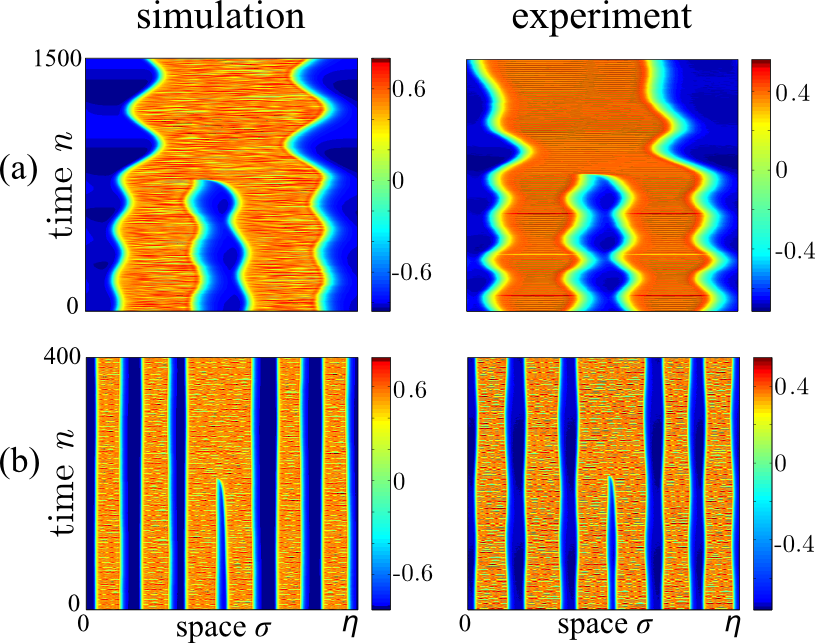}
    \caption{Typical bifurcation events as $N-$headed Chimera becomes
      unstable, being replaced by $(N-1)$ heads. (a): 2-to-1
      transition (point B in Fig.\ref{fig:eps_del}). (b): 6-to-5
      transition (point A in Fig.\ref{fig:eps_del}).}
    \label{fig:bifNtoNm1}
  \end{center}
\end{figure}

Figure \ref{fig:high_order_chim} illustrates how a maximum number
$N_{\sigma}$ is experienced, both in numerics and experiments: for a
fixed parameter setting $(\varepsilon,\delta)$ (point C in
Fig.\ref{fig:eps_del}) corresponding to $N=7$, one observes that an
initial condition imposed as a small sine modulation with 7 periods
within $\eta$, indeed leads to the birth of a 7-headed Chimera. Trying
a small sine modulation with 8 periods results in spontaneous switch
to the birth of a lower number of heads, e.g. $N_\sigma=1$ in the
presented case (but any other $N_\sigma\le N=7$ can be observed in
principle, even chaotic breather, depending on initial conditions).\\
\noindent Looking more carefully at the numerical bifurcation lines in
Fig.\ref{fig:eps_del}, one notices their threshold-like shape, with a
nearly horizontal proportional part close to the origin, and then an
almost vertical part after some threshold. We anticipate that this
peculiar double shape of the bifurcation curves is related to
different bifurcation scenarios for the destabilization of the
$N-$headed Chimera states. Typical $N$ to $(N-1)$ bifurcation events
occurring while crossing such curves, are illustrated in
Fig.\ref{fig:bifNtoNm1}, with captured space-time snapshots for 2-to-1
and 6-to-5 heads transitions. Again, excellent qualitative agreement
is found between experiments and numerics. More detailed analysis of
this novel phenomenon is left to a later report.

\begin{figure}[ht]
  \begin{center}
    \includegraphics[width=0.45\textwidth]{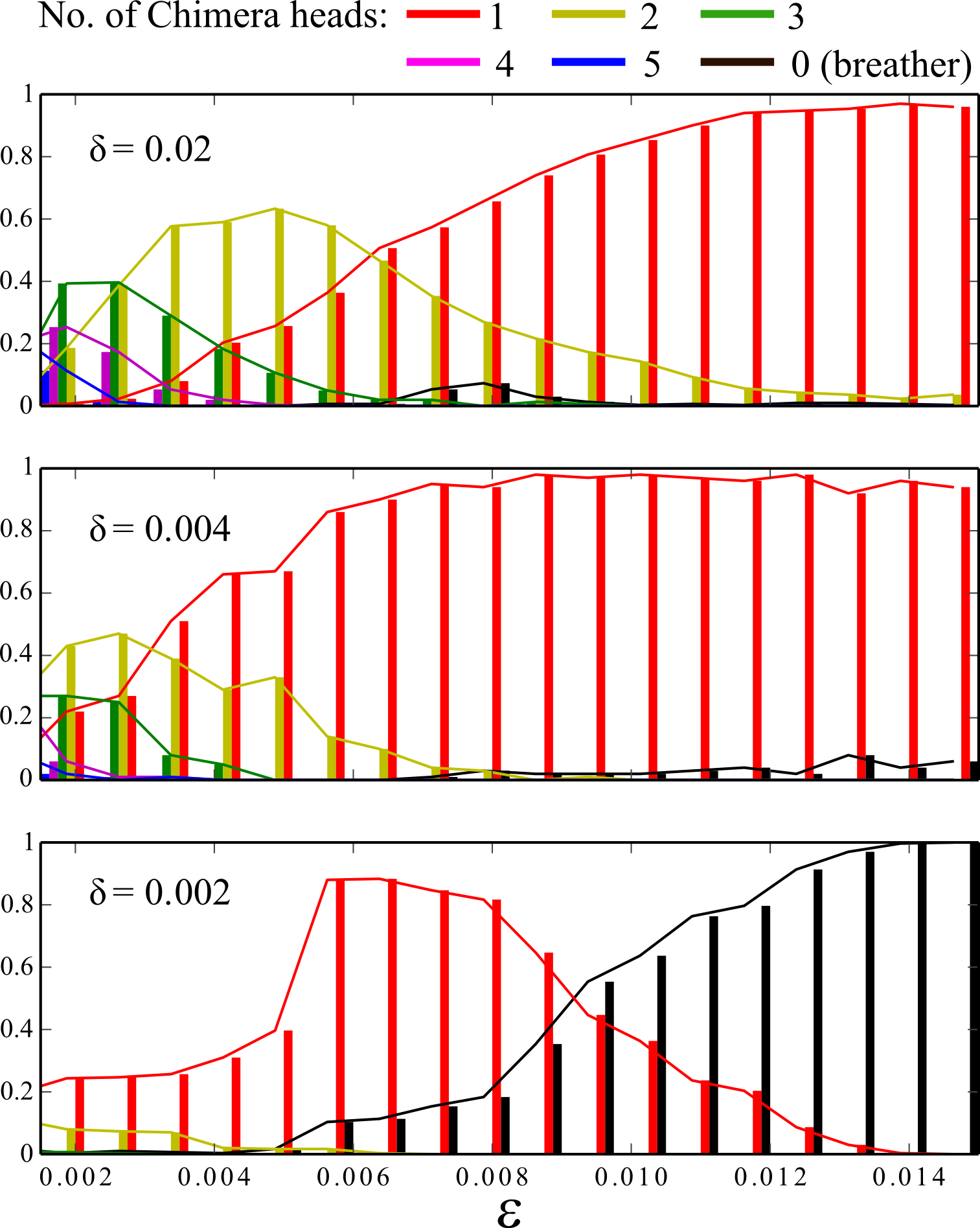}
    \caption{Probability of occurrence under random initial
      conditions, for Chimeras with different number of
      heads \label{fig:basin_stat}, and for three different values of
      $\delta$ (horizontal cut in Fig.\ref{fig:eps_del}).}
  \end{center}
\end{figure}

Figure \ref{fig:eps_del}(a) reveals that model (\ref{eq:idde}) is
highly multistable for small and intermediate $\varepsilon$. This
suggests to explore the basins of attraction of the different
co-existing attractors. Since time-delayed systems are infinite
dimensional through their initial conditions being functional
$\{x_0(s)|\, s\in[0,1]\}$, a precise topological characterization of
the basins structure is not directly possible. One can however try to
estimate the relative size (measure) of the basins in terms of
occurrence probability for each possible solution, after re-setting
many different random initial conditions. This is illustrated in
Fig.\ref{fig:basin_stat}, which shows the evolution vs. $\varepsilon$
of the probability occurrence for $N_\sigma-$headed Chimera for 3
fixed $\delta-$values, with $N_\sigma=0$ to 5 (0 corresponding to
chaotic breather). Each probability has been calculated with 300
different initial noisy conditions $x_0(s)$ (uniform amplitude distribution of $x\in[-1;1]$).\\
In Fig.\ref{fig:basin_stat}(a) ($\delta=0.02$), for small
$\varepsilon$ the most probable solutions are high order multi-headed
Chimera, a small fraction only of initial conditions leading to one-
or two-headed Chimeras. As $\varepsilon$ is increased, $N_\sigma=1$
and $N_\sigma=2$ basins are revealing higher and higher occupation of
phase space. For an intermediate range of $\varepsilon$, one notices
that two-headed Chimera basins reaches a maximum, prevailing on the
other possible $N_\sigma$ with approximately 60\% of probability of
occurrence around $\varepsilon=0.005$. For larger values of
$\varepsilon$, one-headed Chimera basin appears to occupy almost the
full explored phase space. In Fig. \ref{fig:basin_stat}(b) and (c)
corresponding to smaller $\delta-$values, qualitatively similar
features are observed, except that higher-order Chimeras are less and
less probable (they would require smaller $\varepsilon$), and lower
$N_\sigma$ orders predominate together with a growing influence of the
chaotic breather (black) as it is visible already in
Fig.\ref{fig:eps_del}(a) from the positions of the bifurcation
curves.\\
Based on these numerical simulations, we summarize that multi-headed
Chimeras represent an essential part of the solutions exhibited by
Eq.(\ref{eq:idde_norm}), higher order Chimeras requiring small
$\varepsilon$ to predominate.

\begin{figure}[ht]
  \begin{center}
    \includegraphics[width=0.45\textwidth]{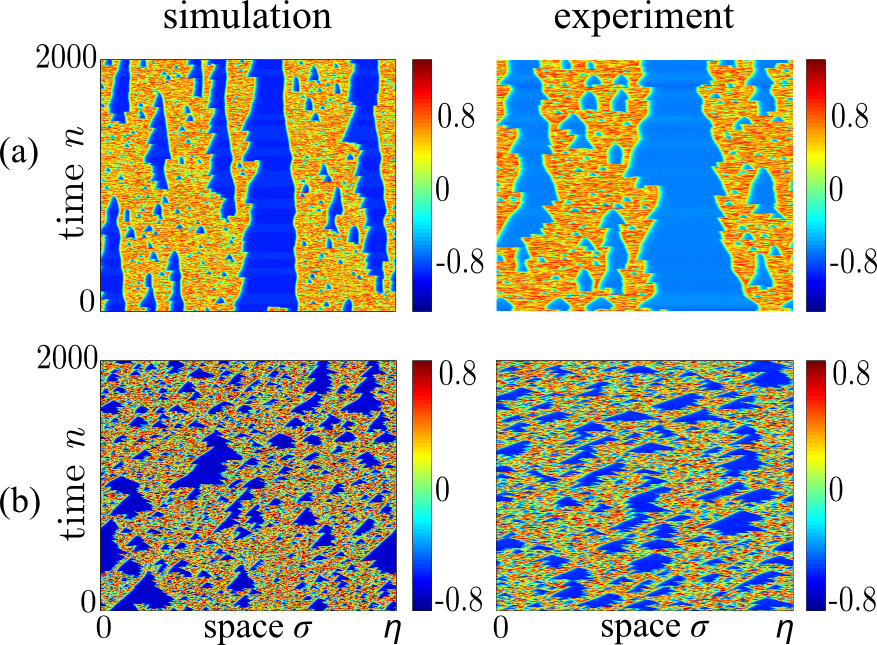}
    \caption{ Progressive route to turbulence as $\beta$
      ($\Phi_0=-0.4$, at point A in the $(\varepsilon,\delta)-$plane
      of Fig.\ref{fig:eps_del}) is increased, revealing more and more
      irregularly vanishing and appearing number of Chimera heads
      $N_\sigma$ as time $n$ is running. (a): $\beta=2.4$. (b):
      $\beta=4.0$.}\label{fig:chim2turb}
  \end{center}
\end{figure}

The bifurcation diagram (Fig. \ref{fig:eps_del}) is obtained for a
fixed normalized gain $\beta=2.0$, providing Chimera states as
alternated chaotic and quite amplitudes.  Increasing $\beta$
progressively destroys the previously sustained Chimera patterns as
$n$ is iterated. Greater $\beta-$values indeed gives rise to a
spatio-temporal turbulent-like intermittent behaviour, including both
chaotic and quite amplitudes in an irregular non-sustained
fashion. This situation is illustrated in Fig.\ref{fig:chim2turb} with
space-time plots for 2 higher values of $\beta$. This transition to
turbulence is again nicely consistent between numerics and
experiments. On the contrary, smaller values of $\beta$ transforms
progressively the chaotic heads into more regular (periodic) or even
quite plateaus, which situation could have been recently analytically
explored \cite{weicker:pre12bis,weicker:ptrsA13}.

\section{Conclusions}

Delay equations have always raised difficult and complex issues,
e.g. motivated whether by technological contexts such as remote
satellite control at the beginning of space exploration, or by the
understanding of complex motion observed in blood cell production
disorders \cite{mackey:sci77}, or even through the quest for optical
chaos \cite{ikeda:oc79}. Beyond theoretical interests, they even led
to surprising experimental research success, e.g. high spectral purity
microwave optoelectronic oscillator for improved Radar performances
\cite{yao:el94}, or demonstration of optical chaos communications for
physical layer encryption in fiber networks \cite{vanwiggeren:sci98,
  goedgebuer:prl98, argyris:nat05}, or even more recently with the
demonstration of novel brain-inspired computing
concepts\cite{jaeger:sci04} in photonic\cite{larger:oe12}. In the
present letter we have reported on a yet unexplored potential of delay
systems in terms of their self-organization capability through a
virtual space-time analogy of delay equations, allowing for the
interpretation of these complex self-organized motions in terms of
Chimera states. The reported excellent agreement between numerical
simulations of a generic model, and the observed phenomena in a
laser-based experiment, suggests the robustness and the intrinsic
character of the underlying dynamical concepts. We anticipate that
such delay systems and their related dynamical phenomena will
represent a simple but efficient theoretical tools for investigation
of complex self-organized motions, as they are naturally developed in
living systems, pattern formation, fluid dynamics phenomena, as well
as behavior in social and technological networks.

This work was supported by the European project PHOCUS (FP7 grant
240763), and the Labex ACTION program (contract ANR-11-LABX-01-01). BP
and YM acknowledge the support of the Region Franche-Comt\'e.

\section*{Methods}
\paragraph*{Experiments:} The acquired signal is corresponding in the
setup of Fig. \ref{fig:lambda_setup}, to the output of the bandpass
filter limiting the dynamics in the feedback loop. This signal
corresponds mathematically to the normalized variable $x(t)$ in
Eqs. (\ref{eq:idde})-(\ref{eq:conv_dde}), and it is physically
proportional to the laser wavelength deviation. As described in the
setup section, an offset current applied to the DBR tuning electrode
of the laser allowed to adjust the normalized parameter $\Phi_0$, thus
allowing the selection of the delayed dynamics operation along a
positive slope of the Fabry-P\'erot Airy function. Chimera pattern can
then be obtained by increasing gradually from zero the feedback loop
gain of the dynamics, which gain is electronically and linearly
adjustable via a DC voltage applied to an analogue electronic
multiplier.\\
A large memory depth (up to 32 million points) digital Lecroy
oscilloscope is used for real-time acquisition of long time traces
covering up to 10000 time delays. This oscilloscope also provides
specific time trace processing capability through short Matlab
routines, thus enabling the real-time visualization of the space-time
patterns shown in Figs. \ref{fig:eps_del}, \ref{fig:high_order_chim},
\ref{fig:bifNtoNm1} and \ref{fig:chim2turb}. The main difficulty for
this custom real time Chimera pattern observation, was to design the
adequate algorithm capable for the accurate ($10^{-5}$ required
precision) and fast extraction of the spatial width $\eta=1+\gamma$,
for which only the Chimera can be clearly observed as a vertical
pattern over thousands of time delay duration. This pattern is whether
strongly tilted in the space-time plane at $10^{-4}$ precision, or
even completely invisible for worth precision. The algorithm aims at
detecting the most frequent time difference between two
plateau-to-chaos sharp transitions in the waveform.  Chimera pattern
evolution were then analyzed while scanning the
$(\varepsilon,\delta)-$plane, which scanning is performed along
hyperbola corresponding to constant characteristic times of the
bandpass filter $\tau$ and $\theta$ (hyperbola equation defined as
$\varepsilon\delta=\tau/\theta=$constant). The hyperbola scan is
obtained through the easy electronic tuning of the time delay $\tau_D$
via the increase of a TTL clock frequency $f_\text{CLK}$. This clock
indeed controls the speed at which the digitized signal is traveling
through the FIFO memory depth used in our digital delay line, the time
delay reads then: $\tau_D=N_0/f_\text{CLK}$ (where $N_0$ is a constant
integer related to the memory depth of 4096, and to the small number
of sampling periods required by the Analog-to-Digital conversion used
in the delay line). Increasing $f_\text{CLK}$ remains to scan the
hyperbola from the top left to the down right, i.e. for decreasing
maximum number of Chimera heads (the highest possible number being
forced by adequate initial conditions, on the top left of the
hyperbola). Different hyperbola were scanned through the choice of
different $\theta$ (i.e. different high pass cut-off frequencies). The
observed Chimera pattern in the graphical Matlab window of the
oscilloscope allowed for the easy detection of the number of heads
$N_\sigma$, or of the chaotic breather solution (0-head), which number
can then reported in Fig. \ref{fig:eps_del}.

\paragraph*{Numerics:} Fourth order Runge-Kutta integration scheme is
used for all the numerical experiments to calculate $x(t)$, with a
fixed time step $h = \varepsilon/10$. The space-time plots are
obtained as in the experiment, through the determination of the
duration $\eta$ revealing ``in average'' vertical patterns, stacking
vertically the $n_0$ successive waveforms $\{x(t) = x_n(\sigma)\, |\,
t=(n\eta+\sigma)\tau_D)\, \text{with}\, \sigma\in[0,\eta],
n=1,2,\ldots,n_0\}$.\\
The procedure for calculating the bifurcation diagram
(Fig. \ref{fig:eps_del}) is the following.  Bifurcation events are
detected for several vertical lines in the
$(\varepsilon,\delta)-$plane (constant $\varepsilon$).  For
progressively decreasing $\delta$, the sustained maximum $N-$headed
Chimera is tested through the calculation of the dynamics over $10^5$
time delay durations, the solution being initially forced (imposing
$x(t)$ for $t\in[-\tau_D;0]$) with a $N-$periods sinusoid. Each
$\delta-$value at which $N$ can not be sustained whereas $N-1$ can, is
used to obtain one point of the $N$ to $N-1$ bifurcation curve.\\
Each chimera basin diagram (Fig. \ref{fig:basin_stat}) is calculated
for $18$ values of $\varepsilon$. The initial interval of $[-\eta;0]$
is interpolated from random initial conditions uniformly distributed
within range $[-1;1]$. For each $\varepsilon$, 300 numerical
experiments is conducted, thus allowing to establish a histogram for
the 300 different asymptotic chimera patterns calculated after $10^5$
time delay evolution.

\section*{Authors' contribution}
LL designed the experiment, performed the measurements, provided the
theoretical dynamics modeling, developed its coupled-network
interpretation, contributed to the analysis, and participated to the
writing. BP performed the numerical simulations and participated to
the experimental record and the writing. YM supervised the Chimera
analysis, provided theoretical interpretations for the observed
phenomena and participated to the writing.

\bibliography{../chimera}

\begin{thebibliography}{10}%
\makeatletter
\providecommand \@ifxundefined [1]{%
 \ifx #1\undefined \expandafter \@firstoftwo
 \else \expandafter \@secondoftwo
\fi
}%
\providecommand \@ifnum [1]{%
 \ifnum #1\expandafter \@firstoftwo
 \else \expandafter \@secondoftwo
\fi
}%
\providecommand \enquote [1]{``#1''}%
\providecommand \bibnamefont  [1]{#1}%
\providecommand \bibfnamefont [1]{#1}%
\providecommand \citenamefont [1]{#1}%
\providecommand\href[0]{\@sanitize\@href}%
\providecommand\@href[1]{\endgroup\@@startlink{#1}\endgroup\@@href}%
\providecommand\@@href[1]{#1\@@endlink}%
\providecommand \@sanitize [0]{\begingroup\catcode`\&12\catcode`\#12\relax}%
\@ifxundefined \pdfoutput {\@firstoftwo}{%
 \@ifnum{\z@=\pdfoutput}{\@firstoftwo}{\@secondoftwo}%
}{%
 \providecommand\@@startlink[1]{\leavevmode\special{html:<a href="#1">}}%
 \providecommand\@@endlink[0]{\special{html:</a>}}%
}{%
 \providecommand\@@startlink[1]{%
  \leavevmode
  \pdfstartlink
   attr{/Border[0 0 1 ]/H/I/C[0 1 1]}%
   user{/Subtype/Link/A<</Type/Action/S/URI/URI(#1)>>}%
  \relax
 }%
 \providecommand\@@endlink[0]{\pdfendlink}%
}%
\providecommand \url  [0]{\begingroup\@sanitize \@url }%
\providecommand \@url [1]{\endgroup\@href {#1}{\urlprefix}}%
\providecommand \urlprefix [0]{URL }%
\providecommand \Eprint[0]{\href }%
\@ifxundefined \urlstyle {%
  \providecommand \doi [1]{doi:\discretionary{}{}{}#1}%
}{%
  \providecommand \doi [0]{doi:\discretionary{}{}{}\begingroup
  \urlstyle{rm}\Url }%
}%
\providecommand \doibase [0]{http://dx.doi.org/}%
\providecommand \Doi[1]{\href{\doibase#1}}%
\providecommand \bibAnnote [3]{%
  \BibitemShut{#1}%
  \begin{quotation}\noindent
    \textsc{Key:}\ #2\\\textsc{Annotation:}\ #3%
  \end{quotation}%
}%
\providecommand \bibAnnoteFile [2]{%
  \IfFileExists{#2}{\bibAnnote {#1} {#2} {\input{#2}}}{}%
}%
\providecommand \typeout [0]{\immediate \write \m@ne }%
\providecommand \selectlanguage [0]{\@gobble}%
\providecommand \bibinfo [0]{\@secondoftwo}%
\providecommand \bibfield [0]{\@secondoftwo}%
\providecommand \translation [1]{[#1]}%
\providecommand \BibitemOpen[0]{}%
\providecommand \bibitemStop [0]{}%
\providecommand \bibitemNoStop [0]{.\EOS\space}%
\providecommand \EOS [0]{\spacefactor3000\relax}%
\providecommand \BibitemShut [1]{\csname bibitem#1\endcsname}%
\bibitem{kuramoto:npcs02}%
  \BibitemOpen
  \bibfield{author}{%
  \bibinfo {author} {\bibfnamefont{Y.}~\bibnamefont{Kuramoto}}\ and\ \bibinfo
  {author} {\bibfnamefont{D.}~\bibnamefont{Battogtokh}},\ }%
  \bibfield{journal}{%
  \bibinfo {journal} {Nonlinear phenomena in complex systems}\ }%
  \textbf{\bibinfo {volume} {5}},\ \bibinfo {pages} {380} (\bibinfo {month}
  {December}\ \bibinfo {year} {2002})%
  \bibAnnoteFile{NoStop}{kuramoto:npcs02}%
\bibitem{abrams:prl04}%
  \BibitemOpen
  \bibfield{author}{%
  \bibinfo {author} {\bibfnamefont{D.~M.}\ \bibnamefont{Abrams}}\ and\ \bibinfo
  {author} {\bibfnamefont{S.~H.}\ \bibnamefont{Strogatz}},\ }%
  \bibfield{journal}{%
  \bibinfo {journal} {Phys. Rev. Lett.}\ }%
  \textbf{\bibinfo {volume} {93}},\ \bibinfo {pages} {174102} (\bibinfo {month}
  {October}\ \bibinfo {year} {2004})%
  \bibAnnoteFile{NoStop}{abrams:prl04}%
\bibitem{hagerstrom:nphy12}%
  \BibitemOpen
  \bibfield{author}{%
  \bibinfo {author} {\bibfnamefont{A.~M.}\ \bibnamefont{Hagerstrom}}, \bibinfo
  {author} {\bibfnamefont{T.~E.}\ \bibnamefont{Murphy}}, \bibinfo {author}
  {\bibfnamefont{R.}~\bibnamefont{Roy}}, \bibinfo {author}
  {\bibfnamefont{P.}~\bibnamefont{H\"ovel}}, \bibinfo {author}
  {\bibfnamefont{I.}~\bibnamefont{Omelchenko}},\ and\ \bibinfo {author}
  {\bibfnamefont{E.}~\bibnamefont{Sch\"oll}},\ }%
  \bibfield{journal}{%
  \bibinfo {journal} {Nature Physics (London)}\ }%
  \textbf{\bibinfo {volume} {8}},\ \bibinfo {pages} {658} (\bibinfo {month}
  {September}\ \bibinfo {year} {2012})%
  \bibAnnoteFile{NoStop}{hagerstrom:nphy12}%
\bibitem{tinsley:nphy12}%
  \BibitemOpen
  \bibfield{author}{%
  \bibinfo {author} {\bibfnamefont{M.}~\bibnamefont{Tinsley}}, \bibinfo
  {author} {\bibfnamefont{S.}~\bibnamefont{Nkomo}},\ and\ \bibinfo {author}
  {\bibfnamefont{K.}~\bibnamefont{Showalter}},\ }%
  \bibfield{journal}{%
  \bibinfo {journal} {Nature Physics (London)}\ }%
  \textbf{\bibinfo {volume} {8}},\ \bibinfo {pages} {662} (\bibinfo {year}
  {2012})%
  \bibAnnoteFile{NoStop}{tinsley:nphy12}%
\bibitem{martens:pnas13}%
  \BibitemOpen
  \bibfield{author}{%
  \bibinfo {author} {\bibfnamefont{E.~A.}\ \bibnamefont{Martens}}, \bibinfo
  {author} {\bibfnamefont{S.}~\bibnamefont{Thutupalli}}, \bibinfo {author}
  {\bibfnamefont{A.}~\bibnamefont{Fourri\`ere}},\ and\ \bibinfo {author}
  {\bibfnamefont{O.}~\bibnamefont{Hallatschek}},\ }%
  \bibfield{journal}{%
  \bibinfo {journal} {Proc. Nat. Acad. Sci.}\ }%
  \textbf{\bibinfo {volume} {110}},\ \bibinfo {pages} {10563–} (\bibinfo
  {month} {June}\ \bibinfo {year} {2013})%
  \bibAnnoteFile{NoStop}{martens:pnas13}%
\bibitem{kapitaniak:sr14}%
  \BibitemOpen
  \bibfield{author}{%
  \bibinfo {author} {\bibfnamefont{T.}~\bibnamefont{Kapitaniak}}, \bibinfo
  {author} {\bibfnamefont{P.}~\bibnamefont{Kuzma}}, \bibinfo {author}
  {\bibfnamefont{J.}~\bibnamefont{Wojewoda}}, \bibinfo {author}
  {\bibfnamefont{K.}~\bibnamefont{Czolczynski}},\ and\ \bibinfo {author}
  {\bibfnamefont{Y.}~\bibnamefont{Maistrenko}},\ }%
  \bibfield{journal}{%
  \bibinfo {journal} {Scient. Rep.}\ }%
  \textbf{\bibinfo {volume} {4}},\ \bibinfo {pages} {6379} (\bibinfo {month}
  {September}\ \bibinfo {year} {2014})%
  \bibAnnoteFile{NoStop}{kapitaniak:sr14}%
\bibitem{larger:prl13}%
  \BibitemOpen
  \bibfield{author}{%
  \bibinfo {author} {\bibfnamefont{L.}~\bibnamefont{Larger}}, \bibinfo {author}
  {\bibfnamefont{B.}~\bibnamefont{Penkovskyi}},\ and\ \bibinfo {author}
  {\bibfnamefont{Y.~L.}\ \bibnamefont{Maistrenko}},\ }%
  \bibfield{journal}{%
  \bibinfo {journal} {Phys. Rev. Lett.}\ }%
  \textbf{\bibinfo {volume} {111}},\ \bibinfo {pages} {054103} (\bibinfo
  {month} {August}\ \bibinfo {year} {2013})%
  \bibAnnoteFile{NoStop}{larger:prl13}%
\bibitem{mackey:sci77}%
  \BibitemOpen
  \bibfield{author}{%
  \bibinfo {author} {\bibfnamefont{M.}~\bibnamefont{Mackey}}\ and\ \bibinfo
  {author} {\bibfnamefont{L.}~\bibnamefont{Glass}},\ }%
  \bibfield{journal}{%
  \bibinfo {journal} {Science}\ }%
  \textbf{\bibinfo {volume} {197}},\ \bibinfo {pages} {287} (\bibinfo {year}
  {1977})%
  \bibAnnoteFile{NoStop}{mackey:sci77}%
\bibitem{ikeda:oc79}%
  \BibitemOpen
  \bibfield{author}{%
  \bibinfo {author} {\bibfnamefont{K.}~\bibnamefont{Ikeda}},\ }%
  \bibfield{journal}{%
  \bibinfo {journal} {Optics Commun.}\ }%
  \textbf{\bibinfo {volume} {30}},\ \bibinfo {pages} {257} (\bibinfo {month}
  {August}\ \bibinfo {year} {1979})%
  \bibAnnoteFile{NoStop}{ikeda:oc79}%
\bibitem{chow:sv87}%
  \BibitemOpen
  \bibfield{author}{%
  \bibinfo {author} {\bibfnamefont{S.}~\bibnamefont{Chow}}\ and\ \bibinfo
  {author} {\bibfnamefont{J.~K.}\ \bibnamefont{Hale}},\ }%
  \enquote{\bibinfo {title} {Dynamics of infinite dimensional systems},}\ \
  (\bibinfo {publisher} {Springer--Verlag},\ \bibinfo {year} {1987})%
  \bibAnnoteFile{NoStop}{chow:sv87}%
\bibitem{leberre:pra90}%
  \BibitemOpen
  \bibfield{author}{%
  \bibinfo {author} {\bibfnamefont{M.}~\bibnamefont{Le~Berre}}, \bibinfo
  {author} {\bibfnamefont{E.}~\bibnamefont{Ressayre}}, \bibinfo {author}
  {\bibfnamefont{A.}~\bibnamefont{Tallet}},\ and\ \bibinfo {author}
  {\bibfnamefont{Y.}~\bibnamefont{Pomeau}},\ }%
  \bibfield{journal}{%
  \bibinfo {journal} {Phys. Rev. A}\ }%
  \textbf{\bibinfo {volume} {41}},\ \bibinfo {pages} {6635} (\bibinfo {month}
  {June}\ \bibinfo {year} {1990})%
  \bibAnnoteFile{NoStop}{leberre:pra90}%
\bibitem{arecchi:pra92}%
  \BibitemOpen
  \bibfield{author}{%
  \bibinfo {author} {\bibfnamefont{F.~T.}\ \bibnamefont{Arecchi}}, \bibinfo
  {author} {\bibfnamefont{G.}~\bibnamefont{Giacomelli}}, \bibinfo {author}
  {\bibfnamefont{A.}~\bibnamefont{Lapucci}},\ and\ \bibinfo {author}
  {\bibfnamefont{R.}~\bibnamefont{Meucci}},\ }%
  \bibfield{journal}{%
  \bibinfo {journal} {Phys. Rev. A}\ }%
  \textbf{\bibinfo {volume} {45}},\ \bibinfo {pages} {R4225} (\bibinfo {month}
  {April}\ \bibinfo {year} {1992})%
  \bibAnnoteFile{NoStop}{arecchi:pra92}%
\bibitem{sharkovsky93}%
  \BibitemOpen
  \bibfield{author}{%
  \bibinfo {author} {\bibfnamefont{A.}~\bibnamefont{Sharkovsky}}, \bibinfo
  {author} {\bibfnamefont{Y.}~\bibnamefont{Maistrenko}},\ and\ \bibinfo
  {author} {\bibfnamefont{E.}~\bibnamefont{Romanenko}},\ }%
  \enquote{\bibinfo {title} {Difference equations and their applications},}\ \
  (\bibinfo {publisher} {Kluwer Acad. Publ. (Naukova Dumka, Kiev, in Russian,
  1986)},\ \bibinfo {year} {1993})\ Chap.~\bibinfo {chapter} {3}%
  \bibAnnoteFile{NoStop}{sharkovsky93}%
\bibitem{giacomelli:epl12}%
  \BibitemOpen
  \bibfield{author}{%
  \bibinfo {author} {\bibfnamefont{G.}~\bibnamefont{Giacomelli}}, \bibinfo
  {author} {\bibfnamefont{F.}~\bibnamefont{Marino}}, \bibinfo {author}
  {\bibfnamefont{M.~A.}\ \bibnamefont{Zaks}},\ and\ \bibinfo {author}
  {\bibfnamefont{S.}~\bibnamefont{Yanchuk}},\ }%
  \bibfield{journal}{%
  \bibinfo {journal} {Europhys. Lett.}\ }%
  \textbf{\bibinfo {volume} {99}},\ \bibinfo {pages} {58005} (\bibinfo {month}
  {September}\ \bibinfo {year} {2012})%
  \bibAnnoteFile{NoStop}{giacomelli:epl12}%
\bibitem{weicker:ptrsA13}%
  \BibitemOpen
  \bibfield{author}{%
  \bibinfo {author} {\bibfnamefont{L.}~\bibnamefont{Weicker}}, \bibinfo
  {author} {\bibfnamefont{T.}~\bibnamefont{Erneux}}, \bibinfo {author}
  {\bibfnamefont{O.}~\bibnamefont{D'Huys}}, \bibinfo {author}
  {\bibfnamefont{J.}~\bibnamefont{Danckaert}}, \bibinfo {author}
  {\bibfnamefont{M.}~\bibnamefont{Jacquot}}, \bibinfo {author}
  {\bibfnamefont{Y.}~\bibnamefont{Chembo}},\ and\ \bibinfo {author}
  {\bibfnamefont{L.}~\bibnamefont{Larger}},\ }%
  \bibfield{journal}{%
  \bibinfo {journal} {Phyl. Trans. Roy. Soc. A}\ }%
  \textbf{\bibinfo {volume} {371}},\ \bibinfo {pages} {20120459} (\bibinfo
  {year} {2013})%
  \bibAnnoteFile{NoStop}{weicker:ptrsA13}%
\bibitem{larger:jqe98}%
  \BibitemOpen
  \bibfield{author}{%
  \bibinfo {author} {\bibfnamefont{L.}~\bibnamefont{Larger}}, \bibinfo {author}
  {\bibfnamefont{J.-P.}\ \bibnamefont{Goedgebuer}},\ and\ \bibinfo {author}
  {\bibfnamefont{J.-M.}\ \bibnamefont{Merolla}},\ }%
  \bibfield{journal}{%
  \bibinfo {journal} {IEEE J. Quantum Electron.}\ }%
  \textbf{\bibinfo {volume} {34}},\ \bibinfo {pages} {594} (\bibinfo {month}
  {April}\ \bibinfo {year} {1998})%
  \bibAnnoteFile{NoStop}{larger:jqe98}%
\bibitem{kouomou:prl05}%
  \BibitemOpen
  \bibfield{author}{%
  \bibinfo {author} {\bibfnamefont{Y.~C.}\ \bibnamefont{Kouomou}}, \bibinfo
  {author} {\bibfnamefont{P.}~\bibnamefont{Colet}}, \bibinfo {author}
  {\bibfnamefont{L.}~\bibnamefont{Larger}},\ and\ \bibinfo {author}
  {\bibfnamefont{N.}~\bibnamefont{Gastaud}},\ }%
  \bibfield{journal}{%
  \bibinfo {journal} {Phys. Rev. Lett.}\ }%
  \textbf{\bibinfo {volume} {95}},\ \bibinfo {pages} {203903} (\bibinfo {month}
  {November}\ \bibinfo {year} {2005})%
  \bibAnnoteFile{NoStop}{kouomou:prl05}%
\bibitem{weicker:pre12bis}%
  \BibitemOpen
  \bibfield{author}{%
  \bibinfo {author} {\bibfnamefont{L.}~\bibnamefont{Weicker}}, \bibinfo
  {author} {\bibfnamefont{T.}~\bibnamefont{Erneux}}, \bibinfo {author}
  {\bibfnamefont{O.}~\bibnamefont{d'Huys}}, \bibinfo {author}
  {\bibfnamefont{J.}~\bibnamefont{Danckaert}}, \bibinfo {author}
  {\bibfnamefont{M.}~\bibnamefont{Jacquot}}, \bibinfo {author}
  {\bibfnamefont{Y.}~\bibnamefont{Chembo}},\ and\ \bibinfo {author}
  {\bibfnamefont{L.}~\bibnamefont{Larger}},\ }%
  \bibfield{journal}{%
  \bibinfo {journal} {Phys. Rev. E}\ }%
  \textbf{\bibinfo {volume} {86}},\ \bibinfo {pages} {055201(R)} (\bibinfo
  {month} {November}\ \bibinfo {year} {2012})%
  \bibAnnoteFile{NoStop}{weicker:pre12bis}%
\bibitem{yao:el94}%
  \BibitemOpen
  \bibfield{author}{%
  \bibinfo {author} {\bibfnamefont{X.~S.}\ \bibnamefont{Yao}}\ and\ \bibinfo
  {author} {\bibfnamefont{L.}~\bibnamefont{Maleki}},\ }%
  \bibfield{journal}{%
  \bibinfo {journal} {Electron. Lett.}\ }%
  \textbf{\bibinfo {volume} {30}},\ \bibinfo {pages} {1525} (\bibinfo {month}
  {September}\ \bibinfo {year} {1994})%
  \bibAnnoteFile{NoStop}{yao:el94}%
\bibitem{vanwiggeren:sci98}%
  \BibitemOpen
  \bibfield{author}{%
  \bibinfo {author} {\bibfnamefont{G.~D.}\ \bibnamefont{VanWiggeren}}\ and\
  \bibinfo {author} {\bibfnamefont{R.}~\bibnamefont{Roy}},\ }%
  \bibfield{journal}{%
  \bibinfo {journal} {Science}\ }%
  \textbf{\bibinfo {volume} {279}},\ \bibinfo {pages} {1198} (\bibinfo {month}
  {February}\ \bibinfo {year} {1998})%
  \bibAnnoteFile{NoStop}{vanwiggeren:sci98}%
\bibitem{goedgebuer:prl98}%
  \BibitemOpen
  \bibfield{author}{%
  \bibinfo {author} {\bibfnamefont{J.-P.}\ \bibnamefont{Goedgebuer}}, \bibinfo
  {author} {\bibfnamefont{L.}~\bibnamefont{Larger}},\ and\ \bibinfo {author}
  {\bibfnamefont{H.}~\bibnamefont{Porte}},\ }%
  \bibfield{journal}{%
  \bibinfo {journal} {Phys. Rev. Lett.}\ }%
  \textbf{\bibinfo {volume} {80}},\ \bibinfo {pages} {2249} (\bibinfo {month}
  {June}\ \bibinfo {year} {1998})%
  \bibAnnoteFile{NoStop}{goedgebuer:prl98}%
\bibitem{argyris:nat05}%
  \BibitemOpen
  \bibfield{author}{%
  \bibinfo {author} {\bibfnamefont{A.}~\bibnamefont{Argyris}}, \bibinfo
  {author} {\bibfnamefont{D.}~\bibnamefont{Syvridis}}, \bibinfo {author}
  {\bibfnamefont{L.}~\bibnamefont{Larger}}, \bibinfo {author}
  {\bibfnamefont{V.}~\bibnamefont{Annovazzi-Lodi}}, \bibinfo {author}
  {\bibfnamefont{P.}~\bibnamefont{Colet}}, \bibinfo {author}
  {\bibfnamefont{I.}~\bibnamefont{Fischer}}, \bibinfo {author}
  {\bibfnamefont{J.}~\bibnamefont{Garcia-Ojalvo}}, \bibinfo {author}
  {\bibfnamefont{C.~R.}\ \bibnamefont{Mirasso}}, \bibinfo {author}
  {\bibfnamefont{L.}~\bibnamefont{Pesquera}},\ and\ \bibinfo {author}
  {\bibfnamefont{A.~K.}\ \bibnamefont{Shore}},\ }%
  \bibfield{journal}{%
  \bibinfo {journal} {Nature (London)}\ }%
  \textbf{\bibinfo {volume} {438}},\ \bibinfo {pages} {343} (\bibinfo {month}
  {November}\ \bibinfo {year} {2005})%
  \bibAnnoteFile{NoStop}{argyris:nat05}%
\bibitem{jaeger:sci04}%
  \BibitemOpen
  \bibfield{author}{%
  \bibinfo {author} {\bibfnamefont{H.}~\bibnamefont{Jaeger}}\ and\ \bibinfo
  {author} {\bibfnamefont{H.}~\bibnamefont{Haas}},\ }%
  \bibfield{journal}{%
  \bibinfo {journal} {Science}\ }%
  \textbf{\bibinfo {volume} {304}},\ \bibinfo {pages} {78} (\bibinfo {month}
  {April}\ \bibinfo {year} {2004})%
  \bibAnnoteFile{NoStop}{jaeger:sci04}%
\bibitem{larger:oe12}%
  \BibitemOpen
  \bibfield{author}{%
  \bibinfo {author} {\bibfnamefont{L.}~\bibnamefont{Larger}}, \bibinfo {author}
  {\bibfnamefont{M.~C.}\ \bibnamefont{Soriano}}, \bibinfo {author}
  {\bibfnamefont{D.}~\bibnamefont{Brunner}}, \bibinfo {author}
  {\bibfnamefont{L.}~\bibnamefont{Appeltant}}, \bibinfo {author}
  {\bibfnamefont{J.~M.}\ \bibnamefont{Gutierrez}}, \bibinfo {author}
  {\bibfnamefont{L.}~\bibnamefont{Pesquera}}, \bibinfo {author}
  {\bibfnamefont{C.~R.}\ \bibnamefont{Mirasso}},\ and\ \bibinfo {author}
  {\bibfnamefont{I.}~\bibnamefont{Fischer}},\ }%
  \bibfield{journal}{%
  \bibinfo {journal} {Opt. Express}\ }%
  \textbf{\bibinfo {volume} {20}},\ \bibinfo {pages} {3241} (\bibinfo {month}
  {January}\ \bibinfo {year} {2012})%
  \bibAnnoteFile{NoStop}{larger:oe12}%
\end{thebibliography}%


\end{document}